\documentclass[aps,twocolumn,prb,showpacs]{revtex4}
\usepackage{graphicx}
\usepackage{dcolumn}
\usepackage{bm}

\newcommand{\nt}{$\nu_T=1$}
\newcommand{\dl}{$d/\ell$}
\newcommand{\dlc}{$(d/\ell)_c$}
\newcommand{\dn}{$\Delta\nu$}

\begin{document}

\title{Charge Imbalance and Bilayer 2D Electron Systems at $\nu_T = 1$}

\author{A.~R. Champagne$^1$, A.~D.~K. Finck$^1$, J.~P. Eisenstein$^1$, L.~N. Pfeiffer$^2$, and K.~W. West$^2$}

\affiliation{$^1$Condensed Matter Physics, California Institute of Technology, Pasadena CA 91125
\\
$^2$Bell Laboratories, Alcatel-Lucent, Murray Hill, NJ 07974}

\date{\today}

\begin{abstract}
We use interlayer tunneling to study bilayer 2D electron systems at \nt\ over a wide range of charge density imbalance, $\Delta \nu =\nu_1-\nu_2$, between the two layers. We find that the strongly enhanced tunneling associated with the coherent excitonic \nt\ phase at small layer separation can survive at least up to an imbalance of \dn\ = 0.5, i.e $(\nu_1, \nu_2)$ = (3/4, 1/4). Phase transitions between the excitonic \nt\ state and bilayer states which lack significant interlayer correlations can be induced in three different ways: by increasing the effective interlayer spacing \dl, the temperature $T$, or the charge imbalance, \dn. We observe that close to the phase boundary the coherent \nt\ phase can be absent at \dn\ = 0, present at intermediate \dn, but then absent again at large \dn, thus indicating an intricate phase competition between it and incoherent quasi-independent layer states. At zero imbalance, the critical \dl\ shifts linearly with temperature, while at \dn\ = 1/3 the critical \dl\ is only weakly dependent on $T$. At \dn\ = 1/3 we report the first observation of a direct phase transition between the coherent excitonic \nt\ bilayer integer quantum Hall phase and the pair of single layer fractional quantized Hall states at $\nu_1$ = 2/3 and $\nu_2=1/3$.
\end{abstract}


\maketitle

\section{Introduction}

A fascinating example of the richness of bilayer 2DES quantum phenomena occurs when the total density $n_T$ of electrons in the bilayer equals the degeneracy $eB/h$ of a single spin-resolved Landau level created by a magnetic field $B$.  In this situation the total Landau level filling factor is $\nu_T = n_T/(eB/h) = 1$. If the spacing between the two layers is small, the 2DES is a gapped quantum Hall (QH) liquid well described by Halperin's $\Psi_{111}$ wavefunction \cite{Halperin83,perspectives} and may be described in several equivalent ways, including as a Bose-Einstein condensate of interlayer excitons\cite{Eisenstein04} or a pseudospin ferromagnet \cite{Yang94,Moon95}. This collective state exists even in the absence of interlayer tunneling\cite{murphy94} and possesses an unusual broken symmetry, spontaneous interlayer phase coherence. This phase coherence is responsible for the very unusual physical properties of the bilayer system at small layer separation, including Josephson-like interlayer tunneling\cite{Spielman00} and vanishing Hall and longitudinal resistances\cite{Kellogg04,Tutuc04,Wiersma04} when currents are driven in opposition (counterflow) in the two layers.  For layer spacings larger than a critical value, interlayer phase coherence is lost and the system properties revert to those characteristic of independent layers. Interlayer tunneling is heavily suppressed at zero bias\cite{jpe92_1}, no anomalous counterflow transport properties are observed and, for density balanced layers (i.e. $\nu_1 = \nu_2 = 1/2$), there is no quantized Hall effect.  For sufficiently large layer separation, the system may be described as two independent composite fermion (CF) metals\cite{jain98,hlr93,shankar03}.

Theory suggests that the transition, as a function of layer separation, between the coherent and incoherent \nt\ states reflects an underlying zero temperature quantum phase transition.  This is consistent with experiments which show that the dimensionless critical layer separation \dlc\ (with $d$ the center-to-center separation of the quantum wells containing the electron gases and $\ell=(\hbar/eB)^{1/2}$ the magnetic length) extrapolates to a finite value $d/\ell \sim 2$ in the $T \rightarrow 0$ limit.
At finite temperatures a Kosterlitz-Thouless (KT) vortex unbinding transition is anticipated\cite{Wen92,Yang94} to lead to the loss of phase coherence above a critical temperature, $T_{KT}$.  Recently, strong experimental evidence for a true finite temperature transition in density balanced \nt\ bilayers has emerged\cite{Champagne08}.  Although the observed linear temperature dependence of the critical layer separation \dlc\ is roughly consistent with theoretical estimates for the KT transition\cite{Moon95}, it remains unclear how close the connection is.  

Spontaneous interlayer phase coherence renders the \nt\ state at small layer separation insensitive to layer density difference, $\Delta\nu = \nu_1-\nu_2$. Indeed, the Halperin $\Psi_{111}$ wavefunction accurately captures the essential correlations at \nt\ and small \dl\ for any combination of densities in the two layers.  Experiments have clearly demonstrated the stability the coherent \nt\ phase against layer density imbalances and have even shown that the critical layer separation \dlc\ is increased by small imbalances\cite{Sawada98,Dolgopolov99,Tutuc03,Spielman04,Clarke05,Wiersma06,Champagne08}.

The robust character of the coherent \nt\ phase at small \dl\ against density imbalance contrasts sharply with the situation at large \dl.  There the bilayer is effectively two independent 2DES layers in parallel and density imbalance opens the possibility of numerous qualitatively different phases.  Conjugate pairs of both incompressible FQHE states, e.g. $(\nu_1,\nu_2)$ = (1/3, 2/3), and compressible states, e.g. $(\nu_1,\nu_2)$ = (1/4, 3/4) are obvious candidates.  But even after specifying the individual filling factors there remain multiple possibilities if the spin polarization of the system is incomplete.  For example, at the low magnetic fields typical of bilayer \nt\ experiments a single layer 2DES at $\nu = 2/3$ is most likely spin unpolarized, even though its conjugate partner at $\nu = 1/3$ is fully polarized.  That spin is a relevant variable in bilayer \nt\ experiments has already been clearly demonstrated\cite{spielman05,kumada05}. 

The above considerations suggest that the phase boundary separating the imbalanced coherent \nt\ bilayer at small \dl\ from the numerous possible incoherent states at large \dl\ is highly intricate.  While disorder within the 2D systems is likely to wash out the finer features of the phase boundary surface, it is reasonable to expect that interesting ones remain.  These features are the subject of this paper.

We report here the results of interlayer tunneling and transport experiments which map out a large portion of the \dl\ - $T$ - \dn\ phase diagram for bilayers at \nt.  We make five main observations: 1) The coherent \nt\ bilayer state can survive at least up to \dn\ = 0.5, i.e. ($\nu_1$, $\nu_2$) = (3/4, 1/4). 2) Deep inside the coherent phase, at small \dl, the effect of a small \dn\ is to reduce the amplitude of the coherent tunneling resonance, while close to the phase boundary, at larger \dl, small imbalances increase the strength of tunneling. 3) Close to the critical $(d/\ell)_c$ the coherent \nt\ phase can be absent at \dn\ = 0, present over an intermediate range of imbalance, and then destroyed again at larger \dn. 4) While the critical layer spacing $(d/\ell)_c$ of density balanced \nt\ bilayers falls linearly with temperature\cite{Champagne08}, at \dn\ = 1/3 we find a much weaker temperature dependence. 5) At \dn\ = 1/3 we find clear evidence that as \dl\ increases the system makes a rapid transition from an incompressible, interlayer phase coherent state to another incompressible, but at most weakly interlayer coupled state consisting of quasi-independent fractional quantized Hall effect (FQHE) states at $\nu_1 = 2/3$ and $\nu_2 = 1/3$. 

The paper is organized as follows: Section II discusses experimental issues, focusing in particular on the challenges encountered in accurately determining the densities of the individual layers in imbalanced bilayers and in making reliable tunneling measurements in systems with low sheet conductivities.  Readers uninterested in these details may wish to skip this section. Section III presents interlayer tunneling and transport data at \nt\ over a broad range of \dn\ , \dl, and $T$. In section IV, we discuss and interpret the data. Section V contains our conclusions.  The Appendix summarizes the numerical modeling used to simulate the interlayer charge transfer effect encountered in the course of these experiments.

\section{Experimental Methods}

\subsection{Sample}

The sample used in this experiment is a GaAs/AlGaAs double quantum well heterostructure grown by molecular beam epitaxy.  Two 18 nm GaAs quantum wells separated by a 10 nm Al$_{0.9}$Ga$_{0.1}$As barrier are embedded between thick cladding layers of Al$_{0.32}$Ga$_{0.68}$As.  Remote Si doping sheets in the cladding layers populate the ground subband of each quantum well with a
2DES with nominal density $5.4 \times 10^{10}$ cm$^{-2}$ and low temperature mobility $1 \times 10^6$ cm$^2$/Vs.  Standard photolithographic techniques are used to confine the bilayer 2DES to a square mesa, 250 $\mu$m on a side.  Four 40 $\mu$m wide arms extend away from the sides of the square to evaporated AuNiGe ohmic contacts.  Each arm is crossed by evaporated metal strip gates on the top and thinned back side of the sample.  These arm gates are used to selectively deplete either the top or bottom 2DES in the bilayer, thereby allowing the ohmic contacts to connect to the inner square region via one or the other 2DES layer separately.  These independent layer contacts are essential for measuring the interlayer tunneling conductance, $dI/dV$ versus interlayer voltage $V$ in the sample.  The densities of the individual 2DESs in the central region of the device are controlled via additional top and back gate voltages, $V_{tg}$ and $V_{bg}$.  Typically the density of each 2DES must be reduced by about a factor of 2 in order to reach the regime where the effective interlayer spacing $d/\ell \sim 1.8$ and interlayer coherence at \nt\ becomes possible.  At these lower densities the mobility of each 2DES is reduced accordingly (in rough proportion to the density) and this in turn adversely affects the strength of correlated electron phenomena in the sample. The problem is amplified when large density imbalances are imposed for then the effects of disorder are asymmetric and can be severe in the low density layer of the pair.

\begin{figure}
\includegraphics[width=3.25in, bb=135 162 387 622]{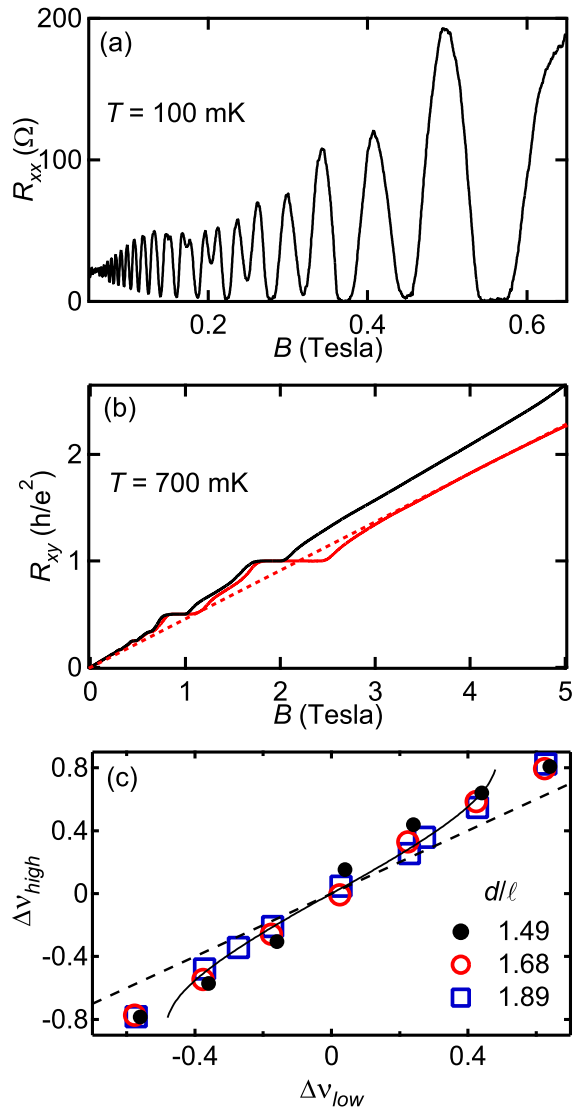}
\caption{\label{}(Color online) (a) Longitudinal magneto-resistance trace in the top 2D gas at low magnetic field $B$ for zero applied top and bottom gate voltages, $V_{tg}$ = $V_{bg}$ = 0, and $T$ = 100 mK. A density of $n_1 = 5.4\times 10^{10}$ cm$^{-2}$ is extracted from the Shubnikov-de Haas oscillations. (b) Hall resistance traces for the bottom 2D gas at $T$ = 700 mK. The carrier density is extracted from the Hall slope. The upper solid (black) trace is acquired with $V_{tg} = -450.3$ mV and $V_{bg}= -3.64$ V.  Under these conditions $n_2 =4.5 \pm 0.15\times10^{10}$ cm$^{-2}$, while the top 2DES is {\it fully depleted}, $n_1=0$.  The lower solid (red) trace is acquired with $V_{tg}=-353.7$ mV and $V_{bg}=-3.33$ V.  With these gate voltages
$n_1 > 0$.  Below $B \approx 1$ T the two traces overlap and hence yield essentially identical values for the lower layer density.  Above $B = 1$ T the traces begin to separate. The dashed line is a linear fit to the lower Hall resistance trace at high fields, giving $n_2 = 5.3 \pm 0.15\times 10^{10}$ cm$^{-2}$. (c) Filling factor imbalance $\Delta \nu_{high}$ deduced from high $B$-field Hall measurements vs. $\Delta \nu_{low}$, the imbalance deduced from low field Shubnikov-de Haas oscillations under the same gating conditions, for three different \dl\ at \nt. The dashed line indicates $\Delta\nu_{high} = \Delta\nu_{low}$.  The solid line is the result of the simulation of the charge transfer effect described in the text and Appendix.}
\end{figure}
\subsection{Imbalanced layers: density calibration}

The density $n$ of a single 2DES is readily determined from conventional magnetotransport measurements.  At low magnetic field or at high temperatures, both of which suppress the quantized Hall effect, the Hall resistance $R_{xy}=B/ne$ is the simplest way to obtain $n$.  Alternatively, the Shubnikov-de Haas oscillations of the longitudinal resistance $R_{xx}$ at intermediate magnetic fields allow for accurate density measurements.  Both of these methods become problematic in the quantized Hall effect regime where plateaus in $R_{xy}$ and wide zeroes in $R_{xx}$ interfere.  

Density determinations in bilayer 2D systems present their own unique challenges, even when independent electrical connections to the individual layers, such as we have here, are available.  At low and intermediate magnetic fields, Hall and Shubnikov-de Haas measurements on the individual layers still provide a satisfactory way to determine the layer densities, $n_1$ and $n_2$.  Figure 1(a) shows typical Shubnikov-de Haas oscillations in the resistivity of the top 2DES.  However, owing to the finite compressibility of the individual 2D electron systems\cite{jpe92_2}, it cannot be assumed that the actions of the top and back-side gates are orthogonal. Instead, it is necessary to independently measure $n_1$ and $n_2$ over an entire grid of gate voltages, $V_{tg}$ and $V_{bg}$.

It is worth noting that interlayer tunneling measurements at zero and low  magnetic fields provide an accurate check of the density calibrations for the special case of equal layer densities. At $B = 0$ the tunneling conductance $dI/dV$ vs. $V$ exhibits a resonance when the subband energies in the two wells line up\cite{jpe91}.  This resonance is centered at $V=0$ only when the densities of the two layers match precisely. It is straightforward to adjust the top and back-side gates to achieve this condition.  Once the bilayer is so balanced, the low field Shubnikov-de Haas-like oscillations in the zero bias tunneling conductance, $G(0)\equiv dI/dV$ at $V=0$, can then be used to extract the layer densities $n_1=n_2$.  

The situation for imbalanced bilayers at high magnetic field is considerably more complicated.  When the top and back-side gates are used to create a density imbalance between the layers, the calibrations discussed above only remain accurate at low and intermediate magnetic fields.  At high fields, in the lowest Landau level, many-body effects lead to a redistribution of charge between the layers. In the following we explain how this effect is observed, understood, and accounted for in the analysis of our experimental results on imbalanced \nt\ bilayers.

Figure 1(b) shows two Hall resistance $R_{xy}$ $vs.$ magnetic field $B$ traces for the bottom 2D electron gas in the bilayer at $T = 700$ mK.  This high temperature was chosen to suppress all but the strongest integer quantum Hall states.  The two traces correspond to two very different gating conditions, but with each designed to produce the same electron density $n_2$ in the bottom 2D layer. For the upper (black) trace the top 2D electron gas is $fully$ depleted, $n_1 =0$, by applying a top gate voltage of $V_{tg}=-450.3$ mV. Meanwhile, the back-side gate voltage is set to $V_{bg}=-3.64$ V, yielding $n_2=4.5\pm 0.15 \times 10^{10}$ cm$^{-2}$.  For the lower (red) trace the gate voltages are $V_{tg}=-353.7$ mV and $V_{bg}=-3.33$ V, yielding $n_2=4.6\pm0.15 \times 10^{10}$ cm$^{-2}$ and $n_1=1.2\pm0.15 \times 10^{10}$ cm$^{-2}$, for the bottom and top 2DESs, respectively.  As expected, at low magnetic fields, $B \lesssim 1$ T, the two Hall traces are essentially identical, demonstrating the near equality of the bottom layer densities in the two cases.  

For $B > 1$ T, corresponding to $\nu_2 < 2$ and the Fermi level of the lower layer entering the lowest orbital Landau level, the two traces begin to separate. For the upper (black) trace, which corresponds to the case where the top 2DES is fully depleted, the slope of the Hall resistance at high magnetic fields gives essentially the same density as found at low fields ($n_2=4.6\pm 0.15 \times 10^{10}$ cm$^{-2}$).  In contrast, in the lower (red) trace, for which the top 2DES is not fully depleted, the Hall slope at high fields is $not$ the same as the slope at low fields. Moreover, the Hall plateaus at $R_{xy}=h/e^2$ and $h/2e^2$ in the lower trace occur at higher magnetic field than in the upper trace.  Apparently, the density of the lower 2DES layer becomes larger at high magnetic fields when a non-zero density of electrons is present in the top layer.  The dashed line is a linear fit to the lower Hall data trace for fields $B > 3.5$ T. This line, which extrapolates back to $R_{xy}=0$ at $B=0$, indicates a density of $n_2=5.3 \pm 0.15  \times 10^{10}$ cm$^{-2}$.  This is substantially larger than the density of $n_2=4.6\pm 0.15 \times 10^{10}$ cm$^{-2}$ inferred from the low field data.  Hall measurements on the top 2DES under comparable gating conditions reveal that its density is $reduced$ as the magnetic field rises above $B = 1$ T.  Our data are consistent, to within experimental error, with the total density $n_T=n_1+n_2$ remaining constant as the magnetic field is increased even though a significant transfer of charge from one layer to the other takes place. 

Figure 1(c) summarizes this density transfer effect for three different total electron densities $n_T$ and thus three different \dl\ values at \nt.  For each \dl\ there are several data points, each corresponding to a specific combination of top and back-side gate voltages, $V_{tg}$ and $V_{bg}$, imposed in order to imbalance the bilayer.  The horizontal coordinate of each data point indicates the filling factor imbalance $\Delta \nu_{low}$ at \nt\ deduced from low field Shubnikov-de Haas measurements made under the same gating conditions.  The vertical coordinate represents the density imbalance $\Delta \nu_{high}$ deduced from the Hall slopes observed at high fields and temperatures, again with same gate voltages.  The dashed line indicates where $\Delta \nu_{low} = \Delta \nu_{high}$; the data do not lie along this line, they merely cross it at \dn\ = 0. The imbalance correction is largest at low \dl\ and large $\Delta\nu$, and measurements have verified that the density changes are equal and opposite in the two layers.

The light solid line in Fig. 1(c) represents a simulation, described in detail in the Appendix, of the charge transfer effect. In essence, the charge transfer is due to the compressibility of the 2D electron systems.  Owing to their finite compressibility, the density of the individual 2DESs in an imbalanced bilayer system does not match the density of positive charges on the gate and in the doping layer closest to it. (Of course, the sum of the two 2DES densities equals the sum of the positive background charge density on both gates and both doping layers.)  If the compressibility is negative\cite{jpe92_2}, the net result is that the imbalance in 2DES densities exceeds the imbalance in the gate charge density.  The solid line in the figure shows that a crude approximation of the effects of negative compressibility captures the essential physics behind the charge transfer effect we observe experimentally.  

In what follows, we refer only to the imbalance \dn\ deduced from high field Hall effect measurements, and thus always incorporate the charge transfer effect.  We stress that for the data presented here the maximum correction to \dn\ never exceeds 0.1 and is generally much smaller.  Indeed, for the majority of the qualitative conclusions we draw, it is immaterial whether the charge transfer effect is included or not.  However, the inclusion of this effect $is$ important in predicting the gate voltages that will render $\nu$ = 1/3 and 2/3 in the individual layers.  We estimate that our quoted imbalances \dn\ are accurate to better than $\pm 0.01$ near \dn\ = 0 and $\pm 0.03$ at $\Delta\nu = 1/3$. 

\subsection{Sheet conductivity versus tunneling conductivity}

The tunneling measurements presented here are two-terminal: a voltage $V$ (consisting of a dc voltage plus a small ($1-10~\mu$V) ac modulation at 3.3 Hz) is applied between the two layers and the resulting ac current is recorded.  The ratio of the ac current to the applied ac modulation voltage yields the conductance $dI/dV$. This conductance is dominated by interlayer tunneling only when the sheet conductivity of the 2D layers themselves is much larger than the tunneling conductance.  If this condition is not met, then only a fraction of the applied voltage drops across the tunnel barrier (with the rest dropping within the 2D layers themselves) and the observed conductance is smaller than the true tunneling conductance\cite{sheetcond1}. 

At high \dl, where the two layers are behaving essentially independently, low sheet conductivity is rarely a problem, at least near zero interlayer voltage.  This is because of the strong suppression of low energy interlayer tunneling at high magnetic field arising from Coulomb blockade-like effects\cite{jpe92_1}. In essence, even if a 2DES is in a thermodynamically compressible state ($e.g.$ at $\nu=1/2$), on the very short time scales characteristic of tunneling the strong $intra$-layer correlations in the system render it effectively incompressible.  The resulting strong suppression of the tunneling conductance allows for reliable measurements even in the presence of relatively low sheet conductivities\cite{sheetcond2}. 

At small \dl, in the interlayer coherent \nt\ phase, a strong and very sharp resonant enhancement of the tunneling conductance $dI/dV$ appears around zero interlayer bias\cite{Spielman00}.  This peak is one of the most dramatic signatures of spontaneous interlayer phase coherence at \nt.  At the same time, its very strength makes tunneling measurements more difficult owing to the finite sheet conductivity of the two 2D layers.  

\begin{figure}
\includegraphics[width=3.25in, bb=169 105 436 475]{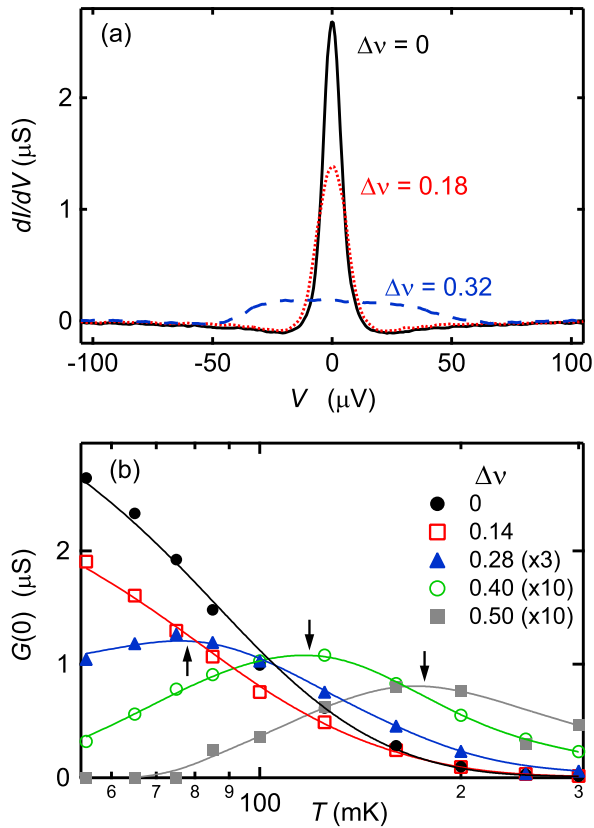}
\caption{\label{} (color online) (a) Tunneling conductance spectra $dI/dV$ vs. $V$ at \nt, $T=55$ mK, and effective layer spacing $d/\ell$ = 1.56, for various charge imbalances, $\Delta\nu =\nu_1-\nu_2$. (b) Temperature dependence of $G(0)$, the height of the tunneling resonance at \nt, for various $\Delta\nu$. The solid lines are guides to the eye. The arrows associated with the \dn\ = 0.28, 0.40 and 0.50 data traces indicate the cross-over temperature $T^*$ defined in the text.}
\end{figure}
Figure 2 illustrates the effects of finite sheet conductivity on tunneling at \nt\ in the coherent phase.  In Fig. 2(a) three \nt\ tunneling conductance resonances are shown. The total electron density in the bilayer ($n_T = 4.95 \times 10^{10}$ cm$^{-2}$), magnetic field ($B=2.05$ T), temperature ($T=55$ mK), and effective layer spacing ($d/\ell = 1.56$) are the same in each case.  At this \dl\ and $T$ the bilayer is relatively deep within the coherent \nt\ phase.  The three resonances differ in the amount of gate-induced density imbalance that is present: \dn\ = 0 for the solid black trace, \dn\ = 0.18 for the short dashed (red) trace, and \dn\ = 0.32 for the long dashed (blue) trace.  For both \dn\ = 0 and \dn\ = 0.18 the tunneling conductance displays the sharp (FWHM $\sim 7 \mu$V) resonance characteristic of the coherent \nt\ phase.  In contrast, at \dn\ = 0.32 the resonance is heavily distorted. This distortion is caused by low sheet conductivity in the low density layer ($n_2 \approx 1.7 \times 10^{10}$ cm$^{-2}$) of the imbalanced bilayer system.  The two-terminal voltage across the sample includes substantial voltage drops within the plane of the low density layer and therefore exceeds the voltage drop across the tunnel barrier.  The tunnel resonance consequently appears to be broadened and its amplitude is reduced. Conventional in-plane transport measurements on the individual layers in the bilayer confirm that their longitudinal conductivity becomes extremely small when their density is reduced sufficiently and the temperature is low.  This is not surprising since electrostatic gating reduces the density of mobile electrons in the 2DES, but does not alter the number of charged impurities in the modulation doping layer.  In effect, gating the density down creates a more disordered 2D electron gas.  At sufficiently low density percolation will cease and the electron gas will become an insulator. As the density is reduced toward this point the sheet conductivity will eventually fall below the interlayer tunneling conductance and tunneling measurements will begin to fail.  Interestingly, in spite of the general decrease of the sheet conductivity in the low density layer, a (distorted) zero bias tunneling peak is still observed, demonstrating that some amount of interlayer coherence remains.  We speculate that this somewhat paradoxical result may be due to strong inhomogeneities within the 2DES, with pockets of interlayer coherent \nt\ fluid surrounded by an incoherent bilayer system in which one layer has very low sheet conductivity.  

It is clearly essential to develop a criterion for distinguishing reliable \nt\ tunneling data for which sheet conductivity effects may be safely ignored from data for which they cannot.  Fortunately, the temperature dependence of the tunneling conductance offers a simple solution to this problem. Figure 2(b) shows the height of the \nt\ tunneling resonance at zero bias, $G(0)$, as a function of temperature for various \dn\ at \dl\ = 1.56. At small imbalance, \dn\ = 0 and 0.14, $G(0)$ increases monotonically as the temperature is reduced and interlayer phase coherence strengthens.  However, at larger \dn, $G(0)$ initially grows as the temperature falls, but then reaches a maximum at some temperature $T^*$ below which the conductance falls with decreasing temperature. For a given imbalance $T^*$ marks the cross-over from a tunneling dominated regime at high temperatures to a sheet conductivity dominated regime at low temperature.  As the data in Fig. 2(b) show, the cross-over temperature $T^*$ is strongly dependent upon \dn.  By acquiring similar data at each \dl\ and \dn\ of interest we determine the minimum temperature at which our conductance data are dominated by interlayer tunneling and are therefore faithful probes of interlayer phase coherence at \nt.  All data presented in the remainder of this paper fulfill this criterion.  

\section{Results}

Figure 3 illustrates the effect of layer charge density imbalance on coherent tunneling at \nt. The figure displays two series of tunneling spectra $dI/dV$ vs. $V$, taken at $T=85$ mK.  The data in panels (a) and (b) were taken at \dl\ = 1.65 and 1.80, respectively. For each trace shown the interlayer voltage $V$ is swept from -150 to +150 $\mu$V.  The traces are offset, by an amount proportional to the imposed imbalance \dn, for clarity. The central trace in each panel is taken in the balanced configuration, \dn\ = 0.  Traces to the upper right of center are increasingly imbalanced by adding charge to the top layer while subtracting it from the bottom layer, yielding $\Delta \nu = \nu_1-\nu_2 >0$.  Traces to the lower left of center correspond to the opposite sense of imbalance: $\Delta \nu <0$.

Figure 3(a) demonstrates that at \dl\ = 1.65 the tunneling peak at $V=0$ is gradually reduced as the bilayer system is increasingly imbalanced.  This result is characteristic of the coherent \nt\ phase at \dl\ well below the critical layer separation\cite{spielmanthesis}.  In contrast, the tunneling data shown in Fig. 3(b) reveal a much more complex dependence upon charge density imbalance.  These data, which were obtained at \dl\ = 1.80 and $T = 85$ mK, show no tunneling peak in the balanced condition \dn\ = 0.  Under these conditions the balanced bilayer system is just outside the coherent \nt\ phase.  As reported previously, imposing a small density imbalance \dn\ creates a zero bias tunneling peak and thus forces the system across the phase boundary and into the coherent phase\cite{Spielman04,Champagne08}. Increasing the imbalance initially strengthens the tunneling peak and drives the system deeper into the coherent phase. Similar conclusions have been reached via conventional magneto-transport measurements\cite{Sawada98,Dolgopolov99,Tutuc03,Clarke05,Wiersma06}.  Here, however, we have the possibility to study the phase boundary out to much larger \dn\ than previously, up to $\Delta \nu \sim 0.4$ for the data in Fig. 3. As Fig. 3(b) demonstrates, the tunneling amplitude reaches a maximum at intermediate imbalance and thereafter is rapidly reduced to zero again.  These results suggest that the critical effective layer separation \dlc\ between the coherent and incoherent phases of \nt\ bilayers may have, in addition to a local minimum at \dn\ = 0, a local $maximum$ at $|\Delta \nu| > 0$.  Additional tunneling and magneto-transport data, to be discussed below, strongly support this conclusion.

\begin{figure}
\includegraphics[width=3.25in, bb=158 78 422 441]{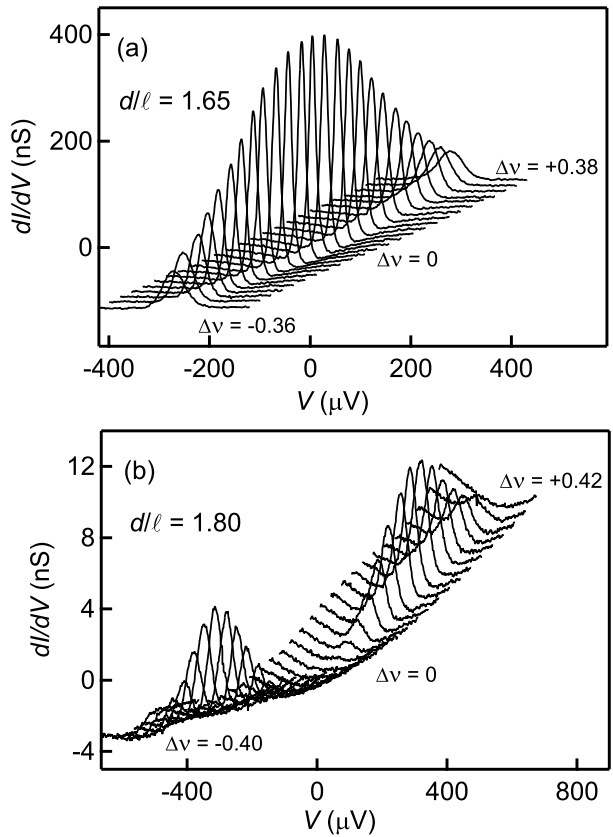}
\caption{\label{} Tunneling conductance spectra $dI/dV$ vs. $V$  at \nt\ and
$T = 85$ mK $vs.$ imbalance \dn\ for (a) $d/\ell$ = 1.65 and (b) $d/\ell$ = 1.80.  In each panel the middle trace corresponds to \dn\ = 0 while the remaining traces are offset by an amount proportional to \dn\ for clarity. The interlayer voltage range is -150 to +150 $\mu$V in all cases.}
\end{figure}
\begin{figure}
\includegraphics[width=3.35in, bb=164 90 434 462]{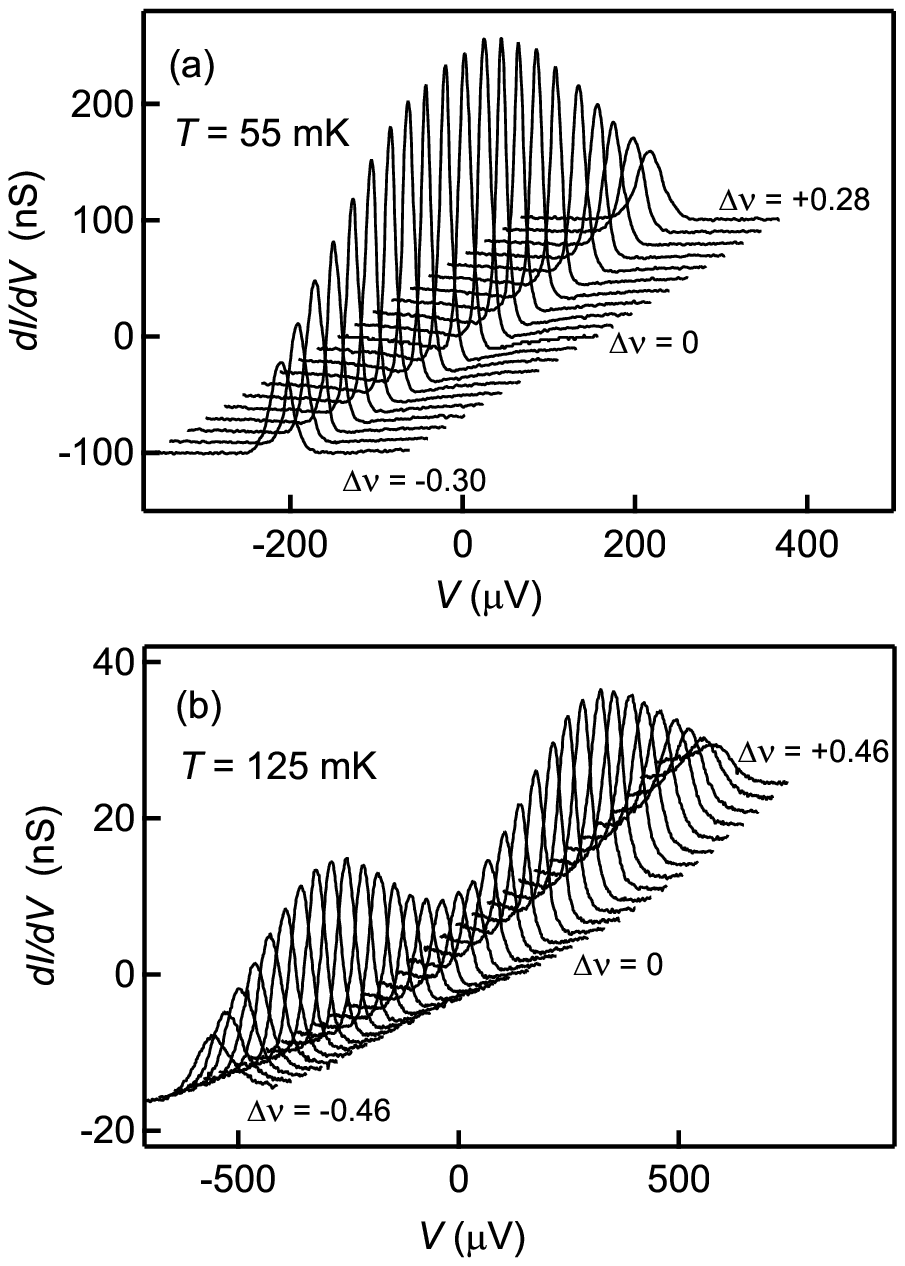}
\caption{\label{} Tunneling conductance spectra $dI/dV$ vs. $V$  at \nt\ and
\dl\ = 1.71 $vs.$ imbalance \dn\ for (a) $T=55$ mK and (b) $T=125$ mK.  In each panel the middle trace corresponds to \dn\ = 0 while the remaining traces are offset by an amount proportional to \dn\ for clarity. The interlayer voltage range is -150 to +150 $\mu$V in all cases.} 
\end{figure}
The data shown in Fig. 3 demonstrate that the imbalance dependence of the tunneling resonance changes significantly when the effective layer separation \dl\ approaches the critical value \dlc.  Recent experiments, however, have shown that \dlc\ itself, at least for balanced \nt\ bilayers, exhibits a significant temperature dependence\cite{Champagne08}.  Thus, if the evolution of the imbalance dependence of the tunneling resonance shown in Fig. 3 signals proximity to the phase boundary, the same qualitative behavior ought to be observable at other temperatures $T$ and effective layer separations \dl\ along that boundary.  In particular, it should be possible to induce the same evolution by changing $T$ at fixed \dl.  Figure 4 demonstrates that this is the case.  In Fig. 4(a) \dl\ = 1.71 and $T= 55$ mK and the system is well inside the coherent phase.  As in Fig. 3(a), under these conditions the strength of the tunneling resonance falls smoothly with increasing imbalance.  In Fig. 4(b) the layer separation is still \dl\ = 1.71 but the temperature has been raised to $T = 125$ mK, thus positioning the system close to the phase boundary.  Here, as in Fig. 3(b), the tunneling resonance initially grows with increasing imbalance, and then falls again at sufficiently large \dn.  Beyond corroborating our earlier evidence\cite{Champagne08} for a finite temperature phase transition in balanced \nt\ bilayers, these new data generalize the notion of a finite temperature phase transition to include highly imbalanced \nt\ bilayers.  The evolution from a local maximum in tunneling at \dn\ = 0 to a local minimum at \dn\ = 0 flanked by local maxima at intermediate \dn\ indicates proximity to a phase transition which can be approached either by raising \dl\ at fixed $T$ or by raising $T$ at fixed \dl.  

Figure 5 summarizes the evolution of the imbalance dependence of the zero bias tunneling conductance $G(0)$ with effective layer separation \dl\ at three different temperatures; $T=85$, 55, and 200 mK in panels (a), (b), and (c), respectively.  At each temperature there is a critical \dlc\ above which $G(0)$ at \dn\ = 0 vanishes, signaling the loss of interlayer phase coherence.  As reported previously\cite{Champagne08}, \dlc\ at \dn\ = 0 falls linearly with increasing temperature.  The data in Fig. 5 also reveal that for \dl\ slightly larger than \dlc\ at \dn\ = 0 there is a range of imbalance centered at \dn\ = 0 over which the \nt\ bilayer remains incoherent. For instance, in Fig. 5(a) the \dl\ = 1.83 data demonstrate a lack of interlayer coherence for $|\Delta\nu| \lesssim 0.11$. Remarkably, for $0.11 \lesssim |\Delta \nu| \lesssim 0.34$ a small, but readily identifiable zero bias tunneling peak is restored.  This strongly suggests that for this effective layer separation, \dl\ = 1.83, the \nt\ bilayer is phase coherent only in this imbalance window.  We emphasize that the collapse of the tunneling resonance at the boundaries of this window is quite rapid.  In particular, near $|\Delta \nu |$ = 0.34 the collapse of the tunneling resonance shown in Fig. 5(a) at \dl\ = 1.83 is far more sudden than the gentle imbalance-induced decline of $G(0)$ observed for $d/\ell \lesssim 1.74$.  Indeed, we believe that the latter effect is not indicative of a phase transition while the former, more sudden, collapse is.  Figures 5(b) and (c) reveal that the same phenomenology observed at $T = 85$ mK and shown in Fig. 5(a) is also encountered at $T = 55$ and 200 mK; the only significant differences are the \dl\ values at which the various features occur. 

A phase boundary for the coherent \nt\ bilayer may be constructed by finding the largest \dl, for a given imbalance \dn\ and temperature $T$, at which a zero bias tunneling resonance is still identifiable above the noise in the tunneling conductance measurement ($\sim 5 \times 10^{-2}$ nS).  Figure 6(a) displays the results of such a construction at $T = 55$, 85, 125, and 200 mK.  At each temperature the coordinates in \dl\ - \dn\ space of the minimum detectable zero bias tunneling resonance are plotted; the associated solid lines are guides to the eye.  As expected, the various curves all display a camel-back shape with a local minimum at the balance point, \dn\ = 0.  In general, the curves move to higher \dl\ values as the temperature is reduced, the effect being most pronounced at the balance point, \dn\ = 0.  Near \dn\ = $\pm 1/3$ the curves cluster together at the lowest $T$.  As we shall discuss, this effect is most likely related to competition between the coherent bilayer \nt\ phase and two quasi-independent single layer fractional quantized Hall states, one at $\nu = 1/3$ and the other at $\nu = 2/3$.

\begin{figure}
\includegraphics[width=3.25in, bb=152 66 419 419]{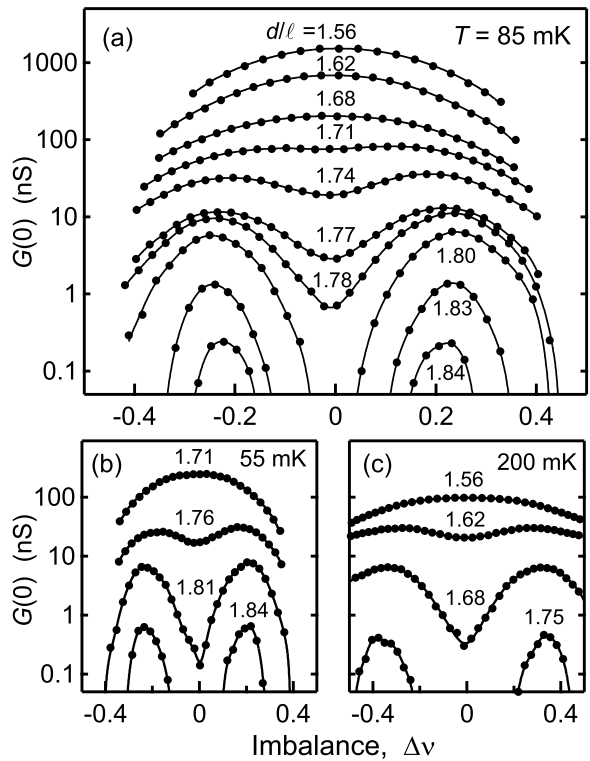}
\caption{\label{} Tunneling conductance at zero bias, $G(0)$, at \nt\ vs. \dn\ for various \dl. (a) $T = 85$ mK, (b) 55 mK, (c) 200 mK. The solid lines are guides to the eye.} 
\end{figure}
Figure 6(b) displays cuts through the phase boundary in the \dl\ - $T$ plane at three different imbalances, \dn\ = 0, 0.16, and 0.33.  The data at \dn\ = 0 (solid circles), represent the phase boundary between the balanced coherent \nt\ bilayer state and a compressible, incoherent state consisting of two quasi-independent $\nu = 1/2$ 2D electron systems.  This phase boundary is linear in temperature over our data range, in agreement with earlier findings\cite{Champagne08} based upon an empirical scaling analyses of the peak tunneling conductance $G(0)$ vs. \dl. At \dn\ = 0.16 the extracted critical layer spacing (open circles) shows a temperature dependence similar to the \dn\ = 0 data, with some hint of saturation at the lowest temperatures.  For the \dn\ = 1/3 data this saturation is pronounced; the critical layer separation (triangles) shows very little temperature dependence for $T \leq 125$ mK.  

\begin{figure}
\includegraphics [width=3.25in, bb=198 129 437 501]{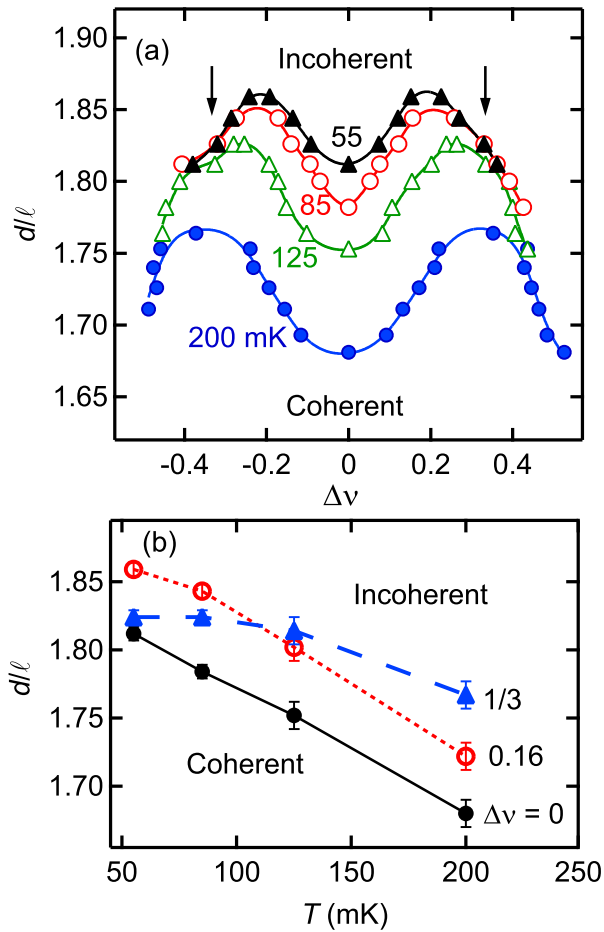}
\caption{\label{} (color online). (a) Phase boundary separating the interlayer coherent \nt\ phase at small $d/\ell$ from the incoherent states at large $d/\ell$ vs. \dn\ at $T$ = 55, 85, 125, and 200 mK. The data points represent the position in $d/\ell$ of the smallest measured tunneling resonance. Arrows denote $\Delta \nu = \pm 1/3$. (b) Phase boundary in the \dl\ - $T$ plane extracted from panel (a) for fixed \dn\ = 0, 0.16, 1/3. The lines are guides to the eye.}
\end{figure}

To investigate in more detail the special case of \dn\ = 1/3, we prepared our bilayer sample with ($\nu_1,\nu_2$) = (2/3, 1/3) and studied both interlayer tunneling and conventional magneto-transport as a function of \dl\ at $T$ = 85 mK. Figure 7(a) shows the tunneling conductance $dI/dV$ vs. $V$ for \dl\ = 1.62, 1.71, 1.80 and 1.84. The strongly enhanced tunneling signature of interlayer phase coherence at \nt\ is clearly visible at low \dl\ but quickly collapses as \dl\ increases, with the tunneling peak disappearing between \dl\ = 1.80 and 1.84. This rapid collapse of the zero bias tunneling resonance with increasing \dl\ is very similar to the one observed for the case of balanced layers where $\nu_1=\nu_2=1/2$ and the bilayer is compressible at large \dl . However, in the present ($\nu_1,\nu_2$) = (2/3, 1/3) situation, the bilayer system at large \dl\ consists of two $incompressible$ FQHE states.  It is therefore plausible that the phase transition we observe is from a coherent \nt\ bilayer whose Hall resistance is quantized at $\rho_{xy}=h/e^2$ to an incoherent system in which one layer exhibits a quantized Hall plateau at $\rho_{xy}=3h/2e^2$ and the other a Hall plateau at $\rho_{xy}=3h/e^2$.  This scenario contrasts starkly with the situation for the balanced \nt\ bilayer.  In that case the Hall resistances of the individual layers at $d/\ell > (d/\ell)_c$ are equal to $\rho_{xy} = 2h/e^2$ and are $not$ quantized.

\begin{figure}
\includegraphics[width=3.25in, bb=173 86 409 453]{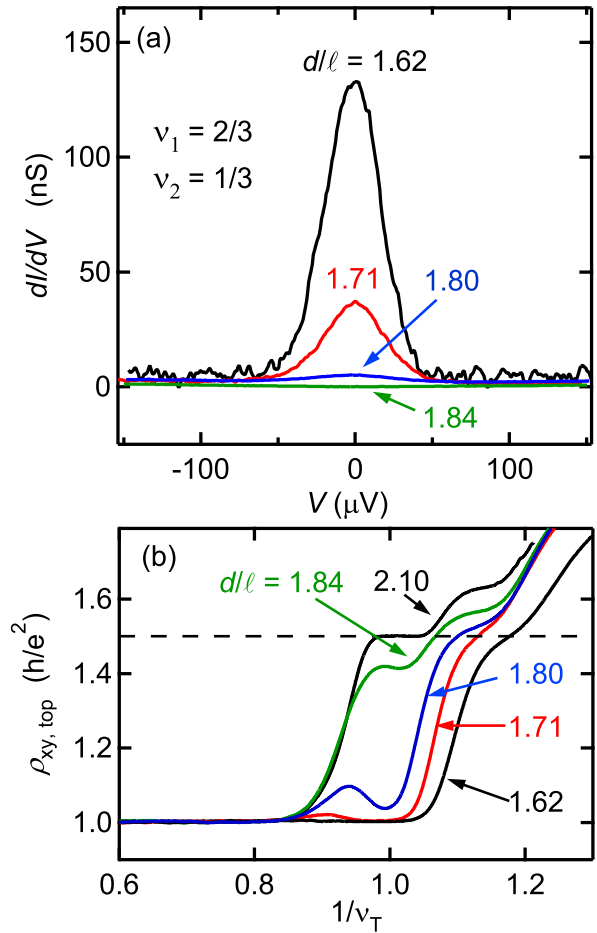} 
\caption{\label{}(color online) Observation of the transition from the interlayer coherent \nt\ phase to two quasi-independent fractional quantized Hall states.  (a) Tunneling conductance spectra at \nt\ and $T$ = 85 mK at fixed imbalance $\Delta \nu = +1/3$ ($i.e.$ $\nu_1=2/3$, $\nu_2=1/3$). Zero bias peak collapses rapidly as \dl\ increases and interlayer coherence is lost. (b) Hall resistance $\rho_{xy,top}$ in the top 2DES $vs.$ inverse {\it total} filling fraction $1/\nu_T$ at $d/\ell$ = 1.62, 1.71, 1.80, 1.84, and 2.10.  For each trace the imbalance at \nt\ is $\Delta \nu = +1/3$. A clear transition from a Hall plateau at $h/e^2$ to one at $3h/2e^2$ is observed at \nt.}
\end{figure}
Figure 7 (b) shows the measured Hall resistance in the top layer as a function of $\nu_T^{-1}$ at various \dl. Deep inside the coherent phase, at \dl\ = 1.62, the Hall resistance at \nt\ exhibits a plateau quantized at $h/e^2$ to within 0.3\%. This result demonstrates that the top layer, whose individual filling factor is $\nu_1 = 2/3$, is part of the coherent \nt\ bilayer state. (It is a remarkable fact that within the coherent \nt\ phase the Hall voltage in either layer is precisely $h/e^2$ times the $total$ current flowing through the bilayer, irrespective of how that current is distributed between the two layers\cite{Kellogg02}.)  At \dl\ = 1.71 the Hall plateau at $h/e^2$ is still present but is narrower than at \dl\ = 1.62.  By \dl\ = 1.80 the transition is imminent; the Hall plateau is gone, being replaced by a local minimum in $\rho_{xy}$ slightly above $h/e^2$. Increasing \dl\ just slightly, to 1.84, has a dramatic effect on $\rho_{xy}$; now a plateau is re-forming, but near $3h/2e^2$.  Further increases of the effective layer spacing lock this nascent plateau accurately onto $\rho_{xy} = 3h/2e^2$, thus proving the existence of the $\nu_1 = 2/3$ FQHE state in the top layer. This rapid transition in $\rho_{xy}$ occurs in the same range of \dl\ values where the zero bias tunneling resonance disappears.  (Similar data reveal that the Hall resistance in the lower layer comes within 3\% of $3h/e^2$, confirming the presence of the 1/3 FQHE state in that layer.)  Taken together, the tunneling and Hall resistance data in Fig. 7 convincingly demonstrate a direct phase transition in imbalanced bilayer \nt\ systems from the interlayer coherent excitonic quantized Hall phase at small \dl\ to quasi-independent fractional quantized Hall states in the individual layers at large \dl.     

\section{Discussion}

The data presented above allow for several new conclusions about the coherent \nt\ bilayer state to be drawn.  Most obviously, the data in Figs. 3, 4 and 5 demonstrate that resonantly enhanced zero bias tunneling is observable in imbalanced \nt\ systems out to $|\Delta \nu | \approx 0.5$. This strongly supports theoretical predictions\cite{perspectives} that interlayer quantum phase coherence, the enabler of collective tunneling, remains intact at these large density imbalances.  

It is nevertheless also clear that large density imbalances reduce the strength $G(0)$ of the zero bias tunneling peak.  Figure 5 shows that this reduction is fairly gentle when $d/\ell < (d/\ell)_c$ and the system is relatively deep within the coherent phase. At larger \dl, closer to the phase boundary, the collapse of $G(0)$ at large \dn\ becomes much more abrupt.  That these abrupt collapses occur at larger \dl, and hence larger total electron density $n_T$  than the more gentle reductions observed at smaller \dl, discounts the possibility that they are caused by disorder in the 2D electron system.  Since the concentration of impurities in the sample is fixed, the effects of disorder are less pronounced at high $n_T$. (This is readily apparent in the mobility of the sample at zero magnetic field.) For this reason we believe that the abrupt collapses of $G(0)$ with \dn\ reflect imbalance-induced phase transitions from coherent to incoherent \nt\ states.  

\begin{figure}
\includegraphics[width=3.25in, bb=0 0 233 158]{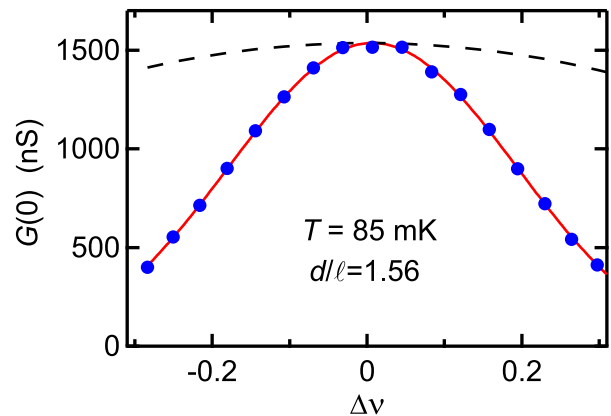}
\caption{\label{} (color online) Tunneling peak height $vs.$ imbalance at $T=85$ mK and \dl\ = 1.56.  The dashed line is prediction of pseudospin moment projection model. The solid line is gaussian fit to the data.} 
\end{figure}
The gentler reductions of $G(0)$ with imbalance observed at smaller \dl\ do not suggest incipient phase transitions and therefore must have a different origin.  
One simple possibility derives from the pseudospin ferromagnetism picture of the coherent \nt\ state.  In this picture an electron definitely in one layer is labeled pseudospin ``up'' while an electron definitely in the other layer is pseudospin ``down''; conventionally these are the eigenstates of the $z$-component of the pseudospin operator, $\tau_z$. In the coherent \nt\ ground state, all electrons are in the $same$ linear combination of up and down states. If the bilayer is density balanced $\langle \tau_z \rangle = \Delta \nu = 0$, and the net pseudospin moment lies in the $x-y$ plane of pseudospin space.  If the bilayer is imbalanced $\langle \tau_z \rangle = \Delta \nu \neq 0$, and the pseudospin moment lies on the surface of a cone symmetric about the $z$ axis.  Relative to the balanced case, the component of the pseudospin moment lying in the $x-y$ plane is reduced by the factor $(1-(\Delta \nu)^2)^{1/2}$.  In an ideal disorder-free sample, an arbitrarily small amount of interlayer tunneling is sufficient to orient the in-plane component of the moment along the $x$-axis.  Since recent theory\cite{Park06} suggests that the zero bias tunneling conductance $G(0)$ is proportional to $\langle \tau_x \rangle^2$, density imbalance should reduce $G(0)$ in direct proportion to $(1-(\Delta\nu)^2)$.  Figure 8 shows a typical comparison between the observed imbalance dependence of $G(0)$ at relatively low \dl\ and this simple moment projection model; clearly the model fails to fit the data.  Near the balance point the observed downward curvature of $G(0)$ vs. \dn\ is about 15 times larger than the model.  The light solid line represents an empirical gaussian fit to the data.

We speculate that the failure of the moment projection model to fit the data may be due to the effects of disorder in the sample.  It has long been suspected that disorder is responsible for the much smaller than expected value of $G(0)$, even in balanced \nt\ bilayers at the lowest \dl\ and temperatures. Disorder almost certainly creates inhomogeneities in the pseudospin field which strongly suppress $\langle \tau_x \rangle^2$ and thereby\cite{Park06} the tunneling conductance $G(0)$. Why this suppression effect apparently becomes more severe as imbalance is imposed remains unclear. 

Figures 3, 4, and 5 also demonstrate that at larger \dl\ the tunneling conductance $G(0)$ exhibits a complex dependence on \dn, including rapid onsets and collapses which we have interpreted as signalling imbalance-driven phase transitions.  Figure 6 summarizes our conclusions about the shape of the phase boundary surface in \dl\ - \dn\ - $T$ space.   The curves shown in Fig. 6(a) were assembled by determining the maximum effective layer separation \dlc\ at which a barely detectable zero bias tunnel resonance could still be observed at a given imbalance \dn\ and temperature $T$.  Figure 6(b) shows the temperature dependence of \dlc\ at \dn\ = 0, 0.16, and 0.33.

The camel-back shape of the phase boundary surface shown in Fig. 6 reflects the interplay of various competing effects.  Near \dn\ = 0 small imbalances increase the critical effective layer separation \dlc\ and thus stabilize the coherent \nt\ phase relative to competing incoherent phases. This result, which has been reported previously\cite{Sawada98,Dolgopolov99,Tutuc03,Spielman04,Clarke05,Wiersma06,Champagne08} was predicted theoretically\cite{Brey,Hanna,Joglekar}.  In their $T=0$ Hartree-Fock analysis Joglekar and MacDonald\cite{Joglekar} found that the collective mode spectrum at \nt\ exhibits a local minimum at finite wavevector analogous to the roton minimum in the excitation spectrum of superfluid helium.  As \dl\ increases this magneto-roton minimum deepens and eventually goes soft at a critical effective layer separation \dlc.  Joglekar and MacDonald identified this point as the phase transition from the interlayer coherent \nt\ quantum Hall phase to a bilayer charge density wave state lacking both interlayer coherence and Hall quantization.  They also found that the collective mode spectrum stiffens when charge density imbalance is imposed and that as a result \dlc\ increases quadratically with \dn. Subsequent experiments\cite{Spielman04} were in qualitative agreement with this prediction.

An intuitive, if oversimplified model\cite{Refael} which explains the initial quadratic increase of \dlc\ with imbalance can be constructed by appealing to a composite fermion (CF) description of the incoherent \nt\ bilayer system.  Close to the critical layer spacing the model assumes that for \dn\ = 0 the difference in the total energies of the coherent and incoherent \nt\ states is $\Delta E = E_{coh}-E_{incoh}=A[d/\ell-(d/\ell)_{c,0}]$, with $A$ a positive constant.  For small imbalances \dn\ the model further assumes that interlayer phase coherence renders any imalance-induced changes in $E_{coh}$ negligible compared to those in $E_{incoh}$. The analogy between a CF metal at $\nu \sim 1/2$ and an ordinary 2D electron gas near zero magnetic field allows for estimating how $E_{incoh}$ depends on \dn. In an imbalanced \nt\ state, the filling factors of the individual layers are $\nu_1=1/2+\Delta \nu /2$ and $\nu_2=1/2-\Delta \nu /2$.  In the composite fermion theory of the half-filled Landau level\cite{hlr93}, deviations of the filling factor from $\nu = 1/2$ correspond to deviations in the effective magnetic field $B_e$ experienced by the CFs from zero; $B_e \propto (\nu-1/2)$ for small $(\nu-1/2)$.  Just as in an ordinary 2DES, when this effective field is weak (and CF Landau levels are not resolved) the total energy of the CF system will increase in proportion to $B_e^2 \propto (\Delta \nu)^2$; this is Landau diamagnetism for composite fermions.  In the imbalanced \nt\ bilayer this diamagnetic effect modifies the energy difference between the coherent and incoherent states: $\Delta E = E_{coh}-E_{incoh}=A[d/\ell-(d/\ell)_{c,0}]-C(\Delta \nu)^2$, with $C$ another positive constant\cite{CFmodel}.  Setting $\Delta E =0$ reveals that the critical layer separation rises quadratically with imbalance, $(d/\ell)_c = (d/\ell)_{c,0}+ C(\Delta \nu)^2/A$, in agreement with experiment.

We now turn to the large \dn\ regime where, as Fig. 6(a) demonstrates, the critical effective layer separation \dlc\ begins to fall with increasing imbalance. One possibility is that the downturn is due to disorder in the sample.  At large imbalance the density of electrons in one of the two layers obviously becomes quite low.  Disorder-induced localization of carriers in that layer might overwhelm the electron-electron interactions responsible for spontaneous interlayer phase coherence.  In this scenario the zero bias tunnel resonance would disappear as \dn\ is increased at fixed \dl, just as Fig. 6(a) suggests.  Increasing the temperature would suppress carrier localization but would also weaken the coherent phase.  Figure 6(b) shows that the net result is complex; at \dn\ = 1/3 there is relatively little effect on \dlc\ until the temperature become quite high.  At very large \dn\ a simple picture might re-emerge.  The high density layer would likely display a traditional $\nu = 1$ integer quantized Hall effect while the low density layer would be an insulator.  Interlayer coherence would presumably be absent.  

Our data clearly demonstrate that disorder-induced localization effects cannot be solely responsible for the shape of the phase boundary at large \dn. Most obviously, the data in Fig. 7 prove that a transition from the interlayer phase coherent \nt\ state to an incoherent pair of quasi-independent FQHE states (with $\nu_1 = 2/3$ and $\nu_2 = 1/3$) has been observed. Interaction effects are essential to the existence of both of these bilayer phases. Figure 6 even contains hints that additional structure may be developing in the phase boundary around \dn\ = $\pm 1/3$ at low temperatures.  These results offer the first proof of phase competition between coherent and incoherent \nt\ imbalanced bilayer states, both of which are incompressible.  To our knowledge, this situation was first examined theoretically, in the context of electron-hole bilayers, by Yoshioka and MacDonald \cite{yoshioka}. 

Figure 6(a) shows that the downturn in the \dlc\ $vs.$ \dn\ phase boundary moves to lower imbalance as the temperature is reduced.  By $T=55$ mK the maxima in  \dlc\ are at $\Delta \nu \approx \pm 0.2$.  This is considerably smaller than $\Delta \nu = \pm 1/3$ where the transition is from the coherent \nt\ phase to a conjugate pair of fractional quantized Hall states.  This observation can be plausibly understood using the same composite fermion description of the incoherent phase that we used previously to intuitively explain the local minimum in the phase boundary at \dn\ = 0.  Initially, as \dn\ grows the total energy of the incoherent phase $E_{incoh}$ rises quadratically with \dn.  At larger \dn\ the CF Landau levels begin to be resolved and oscillations in $E_{incoh}$ result.  This is exactly analogous to the situation with a conventional non-interacting 2D electron gas in a quantizing magnetic field. In that case the Landau splittings are associated with the integer quantized Hall effect; in the present case the CF Landau splittings lead to the fractional QHE. The oscillations in $E_{incoh}$ eventually become quite strong; in an ordinary 2D electron gas the total energy falls all the way back to its zero field value when the Fermi level lies halfway between well-resolved Landau levels.  Although disorder and finite temperature are likely to smear out all but the strongest CF Landau level splittings ($i.e.$ the ones responsible for the $\nu = 1/3$ and 2/3 FQHE states), it is reasonable to expect that $E_{incoh}$ exhibits a maximum somewhere between \dn\ = 0 and \dn\ = 1/3.  If, as assumed previously, the imbalance dependence of the coherent phase is relatively weak, this maximum in $E_{incoh}$ explains the maxima in the phase boundary near $\Delta \nu \approx \pm 0.2$ shown in Fig. 6(a).  

We note in passing that the best current estimates\cite{Rezayi} of the total energy (per electron) of isolated $\nu = 1/2$, 1/3, and 2/3 layers are $-0.466$, $-0.410$, and $-0.518$, respectively\cite{E23}, in units of $e^2/\epsilon \ell$.  From these numbers we conclude that for the same total electron density, two independent 2DES layers have significantly lower energy, by $0.016~e^2/\epsilon \ell$ per electron, when in the imbalanced (1/3, 2/3) state as opposed to the balanced (1/2, 1/2) state\cite{energy}. To the extent that the incoherent \nt\ state in our bilayer samples consists of quasi-independent layers and that the total energy of the coherent phase is only weakly dependent on \dn, this numerical result is certainly consistent with our observation of local maxima in \dlc\ between \dn\ = 0 and \dn\ = $\pm 1/3$ and suggests that the critical layer separation \dlc\ at \dn\ = $\pm 1/3$ might ultimately be found to be even $smaller$ than it is at \dn\ = 0. 

At $|\Delta \nu| > 1/3$ the results in Fig. 6(a) suggest that the critical layer separation continues to fall.  This observation is not understood at present.  Disorder of course becomes more and more important in this regime where the density in one of the two layers becomes very small.  More interesting scenarios including, for example, a Wigner crystal in one layer and a conventional $\nu = 1$ integer QHE in the other, or an exotic excitonic crystalline phase like that suggested by Yang\cite{Yang01}, are also possible.

\section{Conclusion}

Interlayer tunneling spectroscopy has been used to study the effect of layer charge density imbalance on strongly correlated bilayer 2D electron systems at total filling factor \nt.  These tunneling measurements have allowed the determination, over a wide range of density imbalance, temperature, and effective interlayer separation, of the shape of the phase boundary surface separating the spontaneously coherent \nt\ state from various incoherent states. Careful attention was paid to subtle interlayer charge transfer effects which complicate the determination of the imposed density imbalances and to sheet resistance effects which can pollute the determination of the tunneling conductance.

These experiments have demonstrated that the main signature of spontaneous interlayer phase coherence at \nt, a sharp resonance in the tunneling conductance at zero interlayer voltage, is stable against charge density imbalance out to at least $\Delta \nu = \nu_1 - \nu_2 = 0.5$.  At small effective layer separation \dl\ the tunneling resonance strength falls smoothly with increasing \dn, albeit at a rate considerably faster than idealized theory would suggest.  

At larger \dl, close to the phase boundary, a much richer dependence is observed.  Sharp onsets and collapses of the tunneling resonance suggest imbalance-induced phase transitions. We find that the coherent phase can be absent over a range of \dn\ about \dn\ = 0, present at intermediate \dn, and then absent again at large \dn.  This is reflected in a camel-back shape of the critical effective layer separation \dlc\ $vs.$ \dn.

Consistent with prior experiments\cite{Champagne08} we find that the transition between the coherent and various incoherent \nt\ bilayer states can be tuned either by increasing \dl\ at fixed temperature $T$ or by increasing $T$ at fixed \dl. For balanced bilayers we find a linear relationship between the critical \dl\ and temperature.  At \dn\ = 1/3 the temperature dependence of the critical \dl\ is much weaker.

Finally, the first observation of a direct transition from the coherent, and incompressible, \nt\ phase and quasi-independent fractional quantized Hall states lacking interlayer coherence is reported.  At \dn\ = 1/3 increasing \dl\ induces both a rapid collapse of the tunneling resonance and simultaneously a transition between distinct quantized plateaus in the Hall resistances of the individual layers. 

We are grateful to S. Das Sarma, A.H. MacDonald, E.H. Rezayi, X.G. Wen, K. Yang, and especially G. Refael for discussions, and to I.B. Spielman for technical help. This work was supported by the NSF under grant DMR-0552270 and the DOE under grant DE-FG03-99ER45766.

\section{Appendix}
The qualitative origin of the charge transfer effect lies in the negative compressibility of 2D electrons in the lowest Landau level.  Imagine that the doping profile in the sample is perfectly balanced, but that inequivalent electric fields are applied by the top and bottom gates, thereby imbalancing the bilayer.  Owing to the finite compressibility of the 2D electron systems, a non-zero electric field penetrates into the barrier between the layers.  This implies that the density of each layer does not precisely match the sum of the charge density on the gate and in the doping layer closest to it, even though the total system obviously obeys charge neutrality.  Remarkably, the exchange energy of 2D electron systems actually drives the compressibility negative at low density\cite{jpe92_2}.  This leads to the non-intuitive result that reducing the density of one 2D layer with its associated gate can actually slightly increase the density of the other, more remote, layer.  (At low magnetic fields this effect is automatically incorporated into our density calibrations via the Shubnikov-de Haas measurements described above.)  The same basic effect occurs at high magnetic field.  Indeed, the quenching of the kinetic energy created by Landau quantization enhances the importance of many-body effects and renders the compressibility even more strongly negative.  The great complexity of the full density dependence of the total energy of a 2DES in the lowest Landau level makes quantitative modelling impossible, but a qualitative picture is easy to obtain.  To estimate the size of the charge transfer effect we have performed two calculations, one for zero magnetic field and the other for \nt.  (These numerical procedures are described in detail elsewhere\cite{jpe92_2}.) For zero magnetic field we determine the density of each 2DES in an imbalanced bilayer system by self-consistently solving the Schroedinger and Poisson equations, using the local density approximation (LDA) to incorporate the effects of exchange and correlation, for a range of gate voltages and doping densities.  At high magnetic field, in the lowest Landau level, we do a similar calculation except the kinetic energy of the electrons is removed and many-body effects are approximated via the ``backbone'' density dependence of the total energy established by Fano and Ortolani\cite{Fano88}.  While this backbone dependence misses subtle features, like the FQHEs and spontaneous interlayer phase coherence, it captures the qualitative trend toward negative compressibility at low density.  The light solid line in Fig. 1(c) shows typical calculated results for imbalanced bilayers at \nt.  This line is constructed by comparing the LDA calculations of the layer densities at $B = 0$ and at the magnetic field where \nt. In common with the experimental data shown, the theoretical curve indicates that the magnitude of the actual imbalance, $|\Delta \nu_{high}|$, at \nt\ exceeds the value $|\Delta \nu_{low}|$ prescribed by the zero (and low) field density difference\cite{jump}.


\begin{references}
\bibitem{sheetcond1}  The difficulties presented by low sheet conductivity are not readily removed by adding additional voltage probes to the tunneling region. Such probes cannot detect the in-plane voltage drops which appear in the bulk of the individual 2D layers.
\bibitem{sheetcond2} If one or both 2D layers enters a strong quantum Hall state then their sheet conductivities are exponentially suppressed and may again become much smaller than the tunneling conductance.
\bibitem{CFmodel} In this simple model the constant $C=2E_F^{CF}/3$, where $E_F^{CF}$ is the Fermi energy of the $\nu=1/2$ CF liquids in each layer at balance.
\bibitem{E23} The total energy for the $\nu = 2/3$ state follows from that of the 1/3 state and the assumption of particle-hole symmetry; see Ref.\cite{Fano88}.  These numbers assume complete spin polarization.
\bibitem{energy}  For the case of two $\nu = 1/2$ layers, the total energy per unit area is $2 \times 0.5 \times (-0.466) = -0.466$, in units of $(e^2/\epsilon \ell) \times (eB/h)$.  Alternatively, with one layer at $\nu = 1/3$ and the other at $\nu = 2/3$, the total energy, in the same units, is $1/3 \times (-0.410) + 2/3 \times (-0.518) = -0.482$.  Hence, the energetic advantage of the imbalanced case is $0.016 (e^2/\epsilon \ell)$, per electron in the bilayer.
\end{references}
\end{document}